\documentclass[preprint2]{aastex}
\shorttitle{Star Formation in the NGC 2024 Region}
\shortauthors{Choi et al.}

\usepackage{times}
\slugcomment{To appear in the Astronomical Journal}

\begin{document}

\fontsize{10}{10.6}\selectfont

\title{Radio Observations of the Star Formation Activities
       in the NGC 2024 FIR 4 Region} 
\author{\sc Minho Choi$^1$, Miju Kang$^{1,2,3}$, and Jeong-Eun Lee$^4$} 
\affil{$^1$ Korea Astronomy and Space Science Institute,
            776 Daedeokdaero, Yuseong, Daejeon 305-348, Korea;
            minho@kasi.re.kr \\
       $^2$ Max-Planck-Institut f{\"u}r Radioastronomie,
            Auf dem H{\"u}gel 69, D-53121 Bonn, Germany \\
       $^3$ Korea University of Science and Technology,
            217 Gajeongro, Yuseong, Daejeon 305-350, Korea \\
       $^4$ School of Space Research,
            Kyung Hee University, Yongin, Gyeonggi 446-701, Korea}
            
\begin{abstract}
Star formation activities in the NGC 2024 FIR 4 region were studied
by imaging centimeter continuum sources and water maser sources
using several archival data sets from the Very Large Array.
The continuum source VLA 9
is elongated in the northwest-southeast direction,
consistent with the FIR 4 bipolar outflow axis,
and has a flat spectrum in the 6.2--3.6 cm interval.
The three water maser spots associated with FIR 4
are also distributed along the outflow axis.
One of the spots is located close to VLA 9,
and another one is close to an X-ray source.
Examinations of the positions of compact objects in this region suggest
that the FIR 4 cloud core contains a single low-mass protostar.
VLA 9 is the best indicator of the protostellar position.
VLA 9 may be a radio thermal jet driven by this protostar,
and it is unlikely
that FIR 4 contains a high-mass young stellar object (YSO).
A methanol 6.7 GHz maser source is located close to VLA 9,
at a distance of about 100 AU.
The FIR 4 protostar must be responsible for the methanol maser action,
which suggests that methanol class II masers
are not necessarily excited by high-mass YSOs.
Also discussed are properties
of other centimeter continuum sources in the field of view
and the water masers associated with FIR 6n.
Some of the continuum sources are radio thermal jets,
and some are magnetically active young stars.
\end{abstract}


\section{INTRODUCTION}

The NGC 2024 molecular ridge is a filamentary cloud
exhibiting various star formation activities
(Mezger et al. 1988; Visser et al. 1998; Matthews et al. 2002),
at  a distance of 415 pc from the Sun (Anthony-Twarog 1982).
The molecular ridge consists of dense cores FIR 1--7,
and they contain the youngest objects in the NGC 2024 region
(Mezger et al. 1992; Wilson et al. 1995; Chandler \& Carlstrom 1996).
Mezger et al. (1988, 1992) suggested
that the FIR 1--7 cores are isothermal protostars
without luminous stellar objects,
but later studies showed
that some of them are associated with young stellar objects (YSOs)
and molecular outflows (Chandler \& Carlstrom 1996).
While the structures and star formation activities of FIR 5 and FIR 6
are known relatively well
(Richer 1990; Richer et al. 1992; Wiesemeyer et al. 1997;
Lai et al. 2002; Alves et al. 2011; Choi et al. 2012b),
the nature of the other FIR cores are not clearly known.
Recent detections of molecular masers and X-ray emission suggest
that FIR 4 is yet another active site of star formation,
showing interesting physical phenomena at small scales
(Minier et al. 2003; Skinner et al. 2003;
Choi et al. 2012a; Green et al. 2012).

The dense core FIR 4 contains a low-mass protostar
and exhibits star formation activities
such as a reflection nebulosity and a molecular outflow
(Moore \& Chandler 1989; Lis et al. 1991;
Moore \& Yamashita 1995; Chandler \& Carlstrom 1996).
By contrast, Minier et al. (2003) detected a CH$_3$OH class II maser source
and suggested that FIR 4 contains an intermediate/high-mass YSO.
It is unclear whether or not the outflow-driving protostar
and the CH$_3$OH-exciting YSO are the same object.
High-resolution imaging is needed
to identify the YSO in the FIR 4 region and to investigate its nature.
Interferometric observations in the radio continuum
can increase our understanding of the FIR 4 region,
but the imaging of compact sources
is hampered by the strong emission from the extended H {\small II} region
(Crutcher et al. 1986; Gaume et al. 1992).
The effects of the extended emission can be reduced
by excluding the visibility data of short baselines
(Rodr{\'\i}guez et al. 2003; Choi et al. 2012b),
but the resulting images contain artifacts introduced by the filtering
and need to be analyzed carefully.

The 1.3 cm H$_2$O maser of low-mass protostars
is a rarely detectable phenomenon (Kang et al. 2013).
Once detected, however, the maser indicates the existence of shocked gas
and can provide important information
about the star formation activities of YSOs driving the shock
(Genzel \& Downes 1977; Elitzur 1992; Furuya et al. 2003).
Particularly, the small source size and high intensity
allow interferometric observations,
which is essential in the investigation of small-scale structures.
The H$_2$O maser associated with FIR 4
was first reported by Choi et al. (2012a).
They used a single-dish telescope, however,
and the large beam size made it difficult to understand the relation
between the H$_2$O maser source and other compact objects in the region.

In this paper, we present the results of
archival observations of the NGC 2024 FIR 4 region
in the centimeter continuum or in the H$_2$O maser line
with the Very Large Array (VLA)
of the National Radio Astronomy Observatory (NRAO).
The archival data sets are described in Section 2.
In Section 3, we report the results of the radio imaging.
In Section 4, we discuss the star-forming activities of FIR 4.
In Section 5, we describe other objects in the field of view.
A summary is given in Section 6.

\section{DATA}

The VLA data sets presented in this paper
were retrieved from the NRAO Data Archive System
in the form of raw telescope data files.
Table 1 gives a summary of the data sets.
The calibration and imaging were done
using the Astronomical Image Processing System software package of NRAO.

\begin{deluxetable}{llcll}
\tabletypesize{\small}
\tablecaption{Summary of the VLA Data Analyzed}%
\tablewidth{0pt}
\tablehead{
\colhead{Project} & \colhead{Observation Date} & \colhead{Configuration}
& \colhead{Data Type} & \colhead{$F_{\rm ph}$\tablenotemark{a}}}%
\startdata
AG 421\tablenotemark{b}
       & 1994 Aug 13 & B   & 2.0, 3.6, and 6.2 cm continuum
       & \phs0.97 Jy at 3.6 cm \\
       & 1994 Oct 24 & C   & 2.0 and 3.6 cm continuum
       & \phs1.00 Jy at 3.6 cm \\
       & 1995 Aug 17 & A   & 2.0, 3.6, and 6.2 cm continuum
       & \phs1.30 Jy at 3.6 cm \\
AF 354 & 1999 Feb 26 & DnC & 1.3 cm H$_2$O line
       & $\sim$1.73 Jy at 1.3 cm \\
AR 465 & 2002 Mar 2, 3, 8 & A   & 3.6 cm continuum
       & \phs0.76 Jy at 3.6 cm \\
AM 749 & 2003 Mar 14 & D   & 1.3 cm H$_2$O line
       & \phs\nodata \\
AM 780 & 2003 Sep 15 & A   & 1.3 cm H$_2$O line
       & \phs0.83 Jy at 1.3 cm \\
\enddata\\
\tablenotetext{a}{Flux density of the phase calibrator (QSO B0539--057),
                  showing the time variability.}
\tablenotetext{b}{The AG 421 data set presented in this paper is
                  a subset of the full AG 421 data set (see Section 2.1.1).}%
\end{deluxetable}

\subsection{Centimeter Continuum}

We searched VLA data sets in the NRAO Data Archive System
for continuum observations of the FIR 4 region
to investigate the nature and structure of the continuum source
associated with the FIR 4 protostar.
Two data sets were found to be useful.

\subsubsection{Project AG 421}

The NRAO observing project AG 421 includes
data from observations made in various array configurations.
Since we are mainly interested in compact sources,
D-array and DnC-array data were ignored,
and data from more extended configurations were considered.
The B-array, C-array, and A-array observations
were made in 1994 August, 1994 October, and 1995 August, respectively.

The NGC 2024 region was observed in the standard continuum modes
of the $C$ band (4.86 GHz or $\lambda$ = 6.2 cm),
$X$ band (8.44 GHz or $\lambda$ = 3.6 cm),
$U$ band (14.94 GHz or $\lambda$ = 2.0 cm),
and $K$ band (22.46 GHz or $\lambda$ = 1.3 cm).
The phase-tracking center was
$\alpha_{1950}$ = 05$^{\rm h}$39$^{\rm m}$14\fs30,
$\delta_{1950}$ = --01\arcdeg55$'$55\farcs0, which corresponds to IRS 2.
The phase was determined
by observing the nearby quasar 0539--057 (QSO B0539--057).
The flux was calibrated
by observing the quasar 0134+329 (3C 48) or 1328+307 (3C 286).
The flux densities of 0539--057, derived from a comparison of amplitudes,
were 0.96/1.05/1.23 Jy at 6.2 cm, 0.97/1.00/1.30 Jy at 3.6 cm,
and 0.94/1.00/1.16 Jy at 2.0 cm for the B-array/C-array/A-array observations,
respectively.
The observations were made with relatively short scans.
A typical on-source integration time in each array configuration
was $\sim$14 minutes for each wavelength band.

Maps were made using a CLEAN algorithm.
To avoid the adverse effects of the strong extended emission
associated with the nearby ionization fronts,
the visibility data of short baselines ($uv <$ 50 k$\lambda$) were excluded.
This filtering was applied to the data of all array configurations.
For the 6.2 cm continuum, the C-array data were not used
because all the baselines were short ($uv <$ 53 k$\lambda$).
For the 2.0 cm continuum, a $uv$ taper of 750 k$\lambda$ was applied
to make the beam size comparable to that of the 3.6 cm continuum.
The 1.3 cm continuum data did not produce a useful map.
With a robust weighting,
the 6.2 cm, 3.6 cm, and 2.0 cm continuum data produced synthesized beams
of 0\farcs51 $\times$ 0\farcs46, 0\farcs33 $\times$ 0\farcs29,
and 0\farcs33 $\times$ 0\farcs29, respectively, in FWHM.

As a result of the short integration time
and the exclusion of short-spacing data,
the images have relatively low qualities
and contain numerous artifacts in a form of spurious intensity peaks.
Detection of sources is limited by confusion with these unreal peaks.
Therefore, we focused on measuring the flux densities
of the sources at known positions (Rodr{\'\i}guez et al. 2003)
and did not try to identify new sources.
The AG 421 data set is useful for constructing the continuum spectra
of relatively strong sources,
and the AR 465 data set (see below) is useful
for studying source positions and structures.

\subsubsection{Project AR 465}

The NRAO observing project AR 465 includes
data from three tracks of A-array observations made in 2002 March.
The NGC 2024 region was observed
in the standard $X$-band continuum mode (8.46 GHz or $\lambda$ = 3.6 cm).
The phase-tracking center was
$\alpha_{2000}$ = 05$^{\rm h}$41$^{\rm m}$44\fs90,
$\delta_{2000}$ = --01\arcdeg55$'$54\farcs0.
Details of the observations and results were presented
by Rodr{\'\i}guez et al. (2003).
The calibration was done in the way
as described by Rodr{\'\i}guez et al. (2003).
The total on-source integration time was 5.8 hr.

The visibility data from the tree tracks were combined,
and a map was made using a CLEAN algorithm.
The visibility data of short baselines ($uv <$ 50 k$\lambda$) were excluded.
With a robust weighting,
the 3.6 cm continuum data produced a synthesized beam
of FWHM = 0\farcs25 $\times$ 0\farcs23.
(The map presented by Rodr{\'\i}guez et al. 2003 was made
using data with $uv >$ 100 k$\lambda$,
and the data from the three tracks were combined in the image space.)

\subsection{Water Maser}

We searched VLA data sets
for H$_2$O maser observations of the FIR 4 region
to investigate the relation between the maser source and the FIR 4 protostar.
Three data sets were found to have pointing centers close to FIR 4.
In these observations, FIR 4 is located near the edge of the field of view.

\subsubsection{Project AF 354}

The NRAO observing project AF 354 includes
data from the DnC-array observations made in 1999 February.
Details of the observations and results
were presented by Furuya et al. (2003).
For the H$_2$O $6_{16} \rightarrow 5_{23}$ line (22.235077 GHz),
the spectral window was set to have a velocity resolution of 0.33 km s$^{-1}$.
The phase-tracking center was
$\alpha_{1950}$ = 05$^{\rm h}$39$^{\rm m}$13\fs35,
$\delta_{1950}$ = --01\arcdeg57$'$19\farcs0.
The angular distance between the phase-tracking center and FIR 4
is $\sim$70$''$ (Figure 1).
For comparison, the antenna primary beam has an FWHM = 122$''$.
The phase calibrator was the quasar 0539--057 (QSO B0539--057).
The flux was calibrated
by setting the flux density of 0539--057 to 1.73 Jy,
which makes the flux scale consistent with that of Furuya et al. (2003).

\begin{figure}[!p]
\epsscale{1}
\plotone{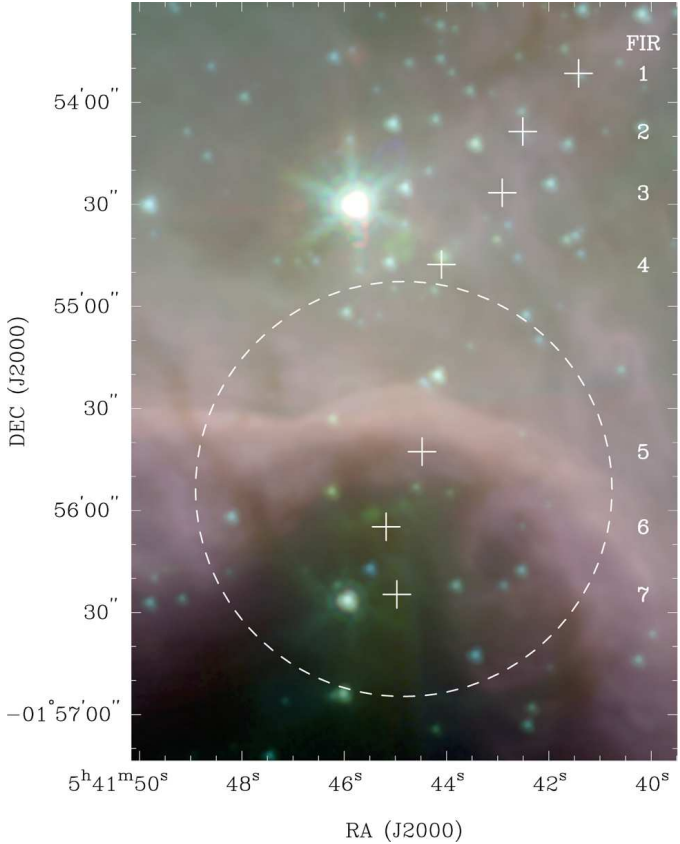}
\caption{
Color composite image
of the 3.6 (blue), 4.5 (green), and 5.8 (red) $\mu$m bands
toward the NGC 2024 region from the {\it Spitzer} data archive.
Plus signs:
870 $\mu$m continuum sources FIR 1--7 (Mezger et al. 1992).
Source names are labeled on the right edge of the figure.
Dashed circle:
FWHM area of the VLA primary beam
centered at the pointing center of the AF 354 observations.
The brightest object in the infrared image is IRS 2 (VLA 19).}
\end{figure}

Maps were made using a CLEAN algorithm.
With a natural weighting,
the H$_2$O line data produced a synthesized beam
of FWHM = 3\farcs1 $\times$ 2\farcs1.
The imaging was done in a square box of 160$''$ $\times$ 160$''$,
which is larger than the size of the primary beam,
in order to include FIR 4 in the imaging area.

\begin{deluxetable}{lrrcrcl}
\tabletypesize{\small}
\tablecaption{NGC 2024 Continuum Source Parameters}%
\tablewidth{0pt}
\tablehead{
Source & \multicolumn{3}{c}{Flux Density\tablenotemark{a}}
& \colhead{$\alpha$\tablenotemark{b}} & \colhead{Associated}
& \colhead{Source} \\
\cline{2-4}
& \colhead{6.2 cm} & \colhead{3.6 cm} & \colhead{2.0 cm}
& & \colhead{Objects} & \colhead{Type\tablenotemark{c}}}%
\startdata
VLA 1  &  0.70 $\pm$ 0.04 &  0.58 $\pm$ 0.03 & OFOV
       & $-$0.33 $\pm$ 0.17 & \nodata & YSO (thermal jet) \\
VLA 2  &  0.47 $\pm$ 0.05 &  0.17 $\pm$ 0.02 & OFOV
       & $-$1.90 $\pm$ 0.33 & \nodata & YSO (magnetic activity) \\
VLA 3  &  6.64 $\pm$ 0.05 &  6.84 $\pm$ 0.03 & \phn4.37 $\pm$ 0.04
       &    0.05 $\pm$ 0.08 & \nodata & YSO (thermal jet) \\
VLA 8  &  2.04 $\pm$ 0.03 &  1.30 $\pm$ 0.02 & \phn0.39 $\pm$ 0.02
       & $-$0.82 $\pm$ 0.09 & \nodata & Background object \\
VLA 9  &  0.51 $\pm$ 0.03 &  0.50 $\pm$ 0.02 & \phn0.97 $\pm$ 0.02
       & $-$0.04 $\pm$ 0.14 & FIR 4   & YSO (thermal jet) \\
VLA 16 &  0.38 $\pm$ 0.03 &  0.58 $\pm$ 0.02 & \nodata
       &    0.76 $\pm$ 0.16 & \nodata & YSO \\
VLA 19 & 13.29 $\pm$ 0.03 & 24.82 $\pm$ 0.02 &    39.92 $\pm$ 0.02
       &    1.13 $\pm$ 0.08 & IRS 2   & Early-B type star \\
VLA 21 &  0.78 $\pm$ 0.03 &  1.56 $\pm$ 0.02 & OFOV
       &    1.26 $\pm$ 0.11 & \nodata & YSO (magnetic activity) \\
VLA 24 &  5.26 $\pm$ 0.03 &  6.06 $\pm$ 0.02 & OFOV
       &    0.26 $\pm$ 0.08 & \nodata & YSO (magnetic activity) \\
VLA 25 &  0.45 $\pm$ 0.04 &  0.62 $\pm$ 0.04 & \nodata
       &    0.59 $\pm$ 0.21 & IRS 4   & YSO (thermal jet) \\
\enddata\\
\tablecomments{See Table 1 of Rodr{\'\i}guez et al. (2003)
               for source positions
               and a more extensive list of associated objects.}%
\tablenotetext{a}{Total flux densities in mJy,
                  corrected for the primary beam response.
                  The uncertainties are the statistical noise of each image.
                  OFOV: located outside the field of view.}%
\tablenotetext{b}{Spectral index in the 6.2--3.6 cm interval:
                  $F_\nu \propto \nu^\alpha$,
                  where $F_\nu$ is the flux density, and $\nu$ the frequency.
                  The uncertainties include the statistical noise
                  and the uncertainty of the flux scale (assumed to be 3\%).}%
\tablenotetext{c}{Nature of the radio source, based on the spectral slope,
                  size, variability, and infrared counterpart.
                  See Section 5.1 for details.}%
\end{deluxetable}

\subsubsection{Project AM 749}

The NRAO observing project AM 749 includes
data from the D-array observations made in 2003 March.
For the H$_2$O line,
the spectral window was set to have a velocity resolution of 1.3 km s$^{-1}$.
The phase-tracking center was
$\alpha_{2000}$ = 05$^{\rm h}$41$^{\rm m}$44\fs765,
$\delta_{2000}$ = --01\arcdeg55$'$53\farcs76.
The angular distance between the phase-tracking center and FIR 4
is $\sim$70$''$.
The phase calibrator was the quasar 0541--056 (QSO B0539--057),
the same as that of AF 354.
The flux was calibrated
by setting the flux density of 0541--056 to 0.828 Jy (see below).

Maps were made using a CLEAN algorithm in the area as described above.
With a natural weighting,
the H$_2$O line data produced a synthesized beam
of FWHM = 4\farcs3 $\times$ 3\farcs4.

\subsubsection{Project AM 780}

The NRAO observing project AM 780 includes
data from the A-array observations made in 2003 September.
The velocity resolution, phase-tracking center, and phase calibrator
were the same as those of AM 749.
The flux was calibrated
by setting the flux density of the quasar 0137+331 (3C 48) to 1.281 Jy.
A comparison of the amplitude
gave a flux density of 0.828 Jy for 0541--056.

Maps were initially made in the area as described above,
which revealed detectable emission near FIR 6n only.
Final maps were made
in a square box of 37$''$ $\times$ 37$''$ around the phase-tracking center.
With a natural weighting,
the H$_2$O line data produced a synthesized beam
of FWHM = 0\farcs13 $\times$ 0\farcs11.

\section{RESULTS}

\subsection{Centimeter Continuum}

Several sources were detected in the centimeter continuum images
of the AG 421 data set.
Table 2 lists the flux densities of the sources detected
in multiple wavelength bands.
VLA 9 is the radio source associated with FIR 4,
and Figure 2 shows the images in the three bands.
The source position agrees with that of the 3.6 cm source
reported by Rodr{\'\i}guez et al. (2003).

\begin{figure*}[!t]
\epsscale{2.0}
\plotone{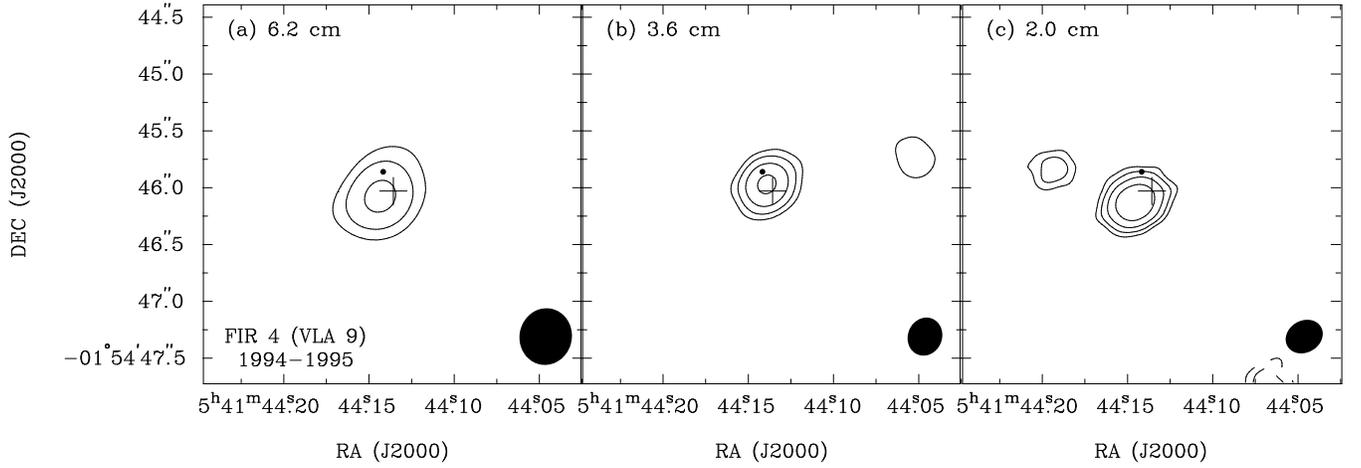}
\caption{
Maps of the 6.2, 3.6, and 2.0 cm continuum emission toward the FIR 4 region,
from the AG 421 data set.
The contour levels are 1, 2, 4, and 8 $\times$ 3$\sigma$,
where $\sigma$ is the rms noise of each image:
0.03, 0.015, and 0.018 mJy beam$^{-1}$, respectively.
Dashed contours are for negative levels.
The maps are corrected for the primary beam response.
Shown in the bottom right-hand corners are the synthesized beams:
FWHM = 0\farcs51 $\times$ 0\farcs46 with P.A. = --11$^\circ$,
FWHM = 0\farcs33 $\times$ 0\farcs29 with P.A. = --21$^\circ$,
and FWHM = 0\farcs33 $\times$ 0\farcs29 with P.A. = --60$^\circ$, respectively.
Plus sign:
3.6 cm continuum source VLA 9 (Figure 3).
Filled circle:
CH$_3$OH 6.7 GHz maser source (Green et al. 2012).
The size of each marker represents
the beam size of the corresponding observations.}
\end{figure*}

\begin{figure}[!b]
\epsscale{1.0}
\plotone{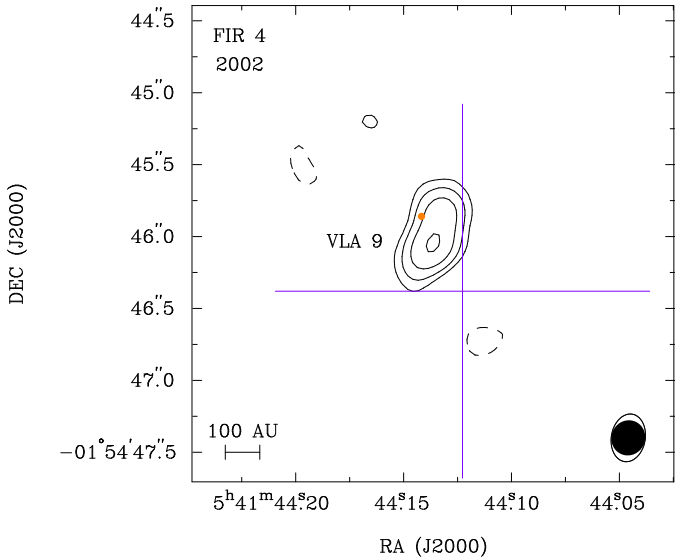}
\caption{
Map of the 3.6 cm continuum emission toward the FIR 4 region,
from the AR 465 data set.
The contour levels are 1, 2, 4, and 8 $\times$ 0.03 mJy beam$^{-1}$,
and the rms noise is 0.011 mJy beam$^{-1}$.
Dashed contours are for negative levels.
The map is corrected for the primary beam response.
Filled ellipse:
synthesized beam.
FWHM = 0\farcs25 $\times$ 0\farcs23 with P.A. = --23$^\circ$.
Open ellipse:
effective beam (apparent width of a point source,
taking the bandwidth smearing into account) at the position of VLA 9,
determined from other unresolved sources in the map.
FWHM = 0\farcs33 $\times$ 0\farcs24 with P.A. = --10$^\circ$.
Large plus sign:
H$_2$O maser spot 1 (Section 3.2).
Orange filled circle:
CH$_3$OH 6.7 GHz maser source (Green et al. 2012).
The size of each marker represents
the beam size of the corresponding observations.
The straight line in the bottom left-hand corner
corresponds to 100 AU at a distance of 415 pc.}
\end{figure}

The 3.6 cm continuum image from the AR 465 observations shows
that VLA 9 has an extended structure (Figure 3).
The peak position is $\alpha_{2000}$ = 05$^{\rm h}$41$^{\rm m}$44\fs137,
$\delta_{2000}$ = --01\arcdeg54$'$46\farcs06,
consistent with Rodr{\'\i}guez et al. (2003).
The total flux density is 0.57 $\pm$ 0.02 mJy.
The source is elongated in the northwest-southeast direction.
The extent of elongation is larger than the effective beam
(synthesized beam degraded owing to the bandwidth smearing).
In addition, the intensity distribution is slightly curved
and asymmetric with respect to the peak position.
An elliptical Gaussian fit gives a deconvolved size
of FWHM = 0\farcs37 $\times$ 0\farcs09
with a position angle (P.A.) = --25$^\circ$
(deconvolved with the effective beam).
The major axis corresponds to FWHM = 150 AU.

The spectrum of VLA 9 in the 6.2--3.6 cm interval is flat (Table 2),
which suggests that the centimeter continuum
is optically thin free-free emission.
This spectrum rules out the possibility of VLA 9
being an ultracompact/hypercompact H {\small II} region.
The elongated structure and the flat spectrum suggest
that VLA 9 is a radio thermal jet.
The 2.0 cm flux density is higher than those of the longer wavelengths,
and the spectral index in the 3.6--2.0 cm interval is $\sim$1.2 (Figure 4),
which suggests that a fraction of the 2.0 cm flux may come from dust.

\begin{figure}[!t]
\epsscale{1.0}
\plotone{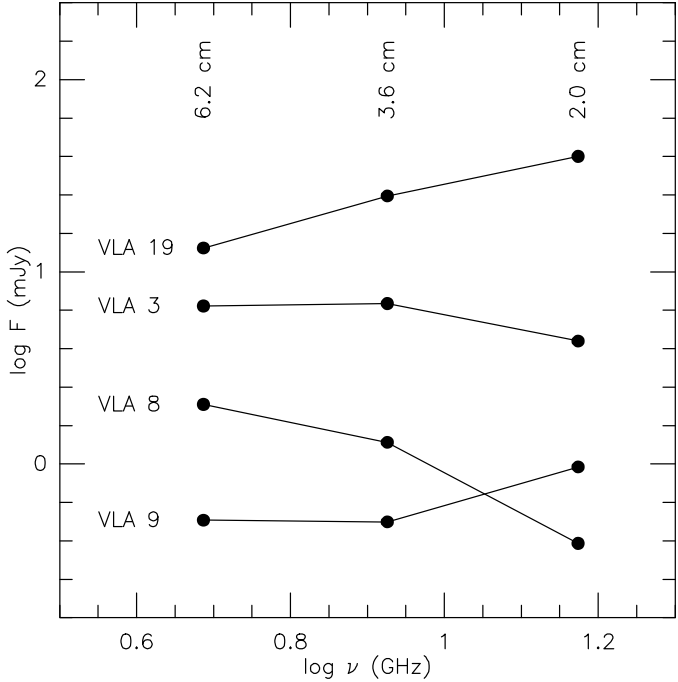}
\caption{
Spectral energy distributions of the sources
detected in the three wavelength bands (Table 2).
The flux uncertainties (including the uncertainty of the flux scale)
are smaller than the sizes of the markers.}
\end{figure}

\subsection{Water Maser}

The image from the AF 354 data set
revealed detectable H$_2$O maser spots near FIR 4 and FIR 6n.
The FIR 6n maser source was already reported by Furuya et al. (2003),
and we will omit descriptions of this source.
(Also see the clarification of source identification in Choi et al. 2012a.)
No maser source was detected from the AM 749 data set.
The 3$\sigma$ detection limits
are 8 mJy beam$^{-1}$ at FIR 6n and 19 mJy beam$^{-1}$ at FIR 4.
The AM 780 data set revealed detectable H$_2$O maser spots near FIR 6n only.
The detection limit at FIR 4 is 40 mJy beam$^{-1}$.
The results are summarized in Table 3.

\begin{deluxetable}{lccllcc}
\tabletypesize{\small}
\tablecaption{\small NGC 2024 FIR 4/6 Water Maser Source Parameters}
\tablewidth{0pt}
\tablehead{
\colhead{Region} & \colhead{Epoch} & \colhead{Spot}
& \multicolumn{2}{c}{Peak Position\tablenotemark{a}}
& \colhead{Centroid Velocity} & \colhead{Peak Flux Density\tablenotemark{b}} \\
\cline{4-5}
&&& \colhead{$\alpha_{\rm J2000.0}$} & \colhead{$\delta_{\rm J2000.0}$}
& \colhead{(km s$^{-1}$)} & \colhead{(Jy beam$^{-1}$)}}
\startdata
FIR 4  & 1999 Feb 26
&  1 & 05 41 44.12  & --01 54 46.4  & 11.3 & 0.16\phn\ $\pm$ 0.04\phn \\
&& 2 & 05 41 44.18  & --01 54 48.9  & 11.4 & 0.10\phn\ $\pm$ 0.04\phn \\
&& 3 & 05 41 44.02  & --01 54 43.3  & 19.1 & 4.29\phn\ $\pm$ 0.04\phn \\
FIR 6n & 2003 Sep 15
&  4 & 05 41 45.155 & --01 56 00.64 & 10.7 & 3.536     $\pm$ 0.009 \\
&& 5 & 05 41 45.171 & --01 56 00.60 & 13.7 & 0.897     $\pm$ 0.009 \\
&&   &              &               & 17.1 & 1.546     $\pm$ 0.009 \\
&&   &              &               & 21.9 & 0.987     $\pm$ 0.009 \\
\enddata\\
\tablenotetext{a}{Units of R.A. are hours, minutes, and seconds,
                  and units of decl. are degrees, arcminutes, and arcseconds.}
\tablenotetext{b}{Flux density at the peak channel,
                  corrected for the primary beam response.}
\end{deluxetable}

Three H$_2$O maser spots associated with FIR 4
were detected from the AF 354 observations:
two weak sources (spots 1 and 2)
near the systemic velocity of the ambient cloud
($V_{\rm LSR}$ = 11.0 km s$^{-1}$; Schulz et al. 1991)
and a strong one (spot 3) at a redshifted velocity (Figures 5 and 6).
The velocity difference between them is $\sim$8 km s$^{-1}$.
The H$_2$O maser source detected
by the single-dish observations in 2012
had a velocity of $V_{\rm LSR}$ = 13 km s$^{-1}$ (Choi et al. 2012a).

\begin{figure}[!t]
\epsscale{1.0}
\plotone{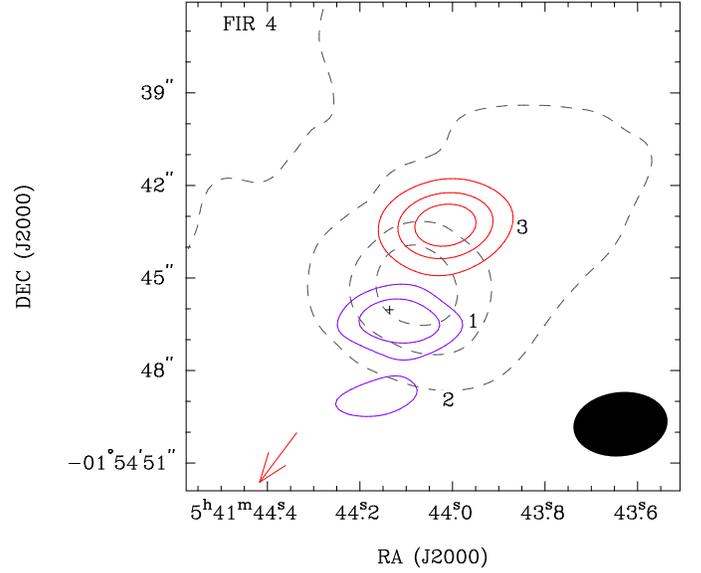}
\caption{
Maps of the H$_2$O maser emission in the FIR 4 region.
The maser spot numbers are labeled.
The violet contours show the intensity
integrated over $V_{\rm LSR}$ = (10.7, 11.7) km s$^{-1}$,
and the contour levels are 0.5 and 0.75 times the peak intensity of spot 1
(0.14 Jy beam$^{-1}$ km s$^{-1}$).
The red contours show the intensity integrated over (18.6, 19.6) km s$^{-1}$,
and the contour levels are 0.25, 0.5, and 0.75 times the peak intensity
(2.67 Jy beam$^{-1}$ km s$^{-1}$).
The maps are corrected for the primary beam response.
Shown in the bottom right-hand corner is the synthesized beam:
FWHM = 3\farcs1 $\times$ 2\farcs1 with a P.A. = --84$^\circ$.
Gray dashed contours:
map of the 4.5 $\mu$m emission from the {\it Spitzer} data archive.
The contour levels are 1, 2, and 4 $\times$ 200 MJy sr$^{-1}$.
Plus sign:
3.6 cm continuum source VLA 9 (Figure 3).
Red arrow:
direction of the redshifted CO outflow (Chandler \& Carlstrom 1996).}
\end{figure}

\begin{figure}[!t]
\epsscale{1.0}
\plotone{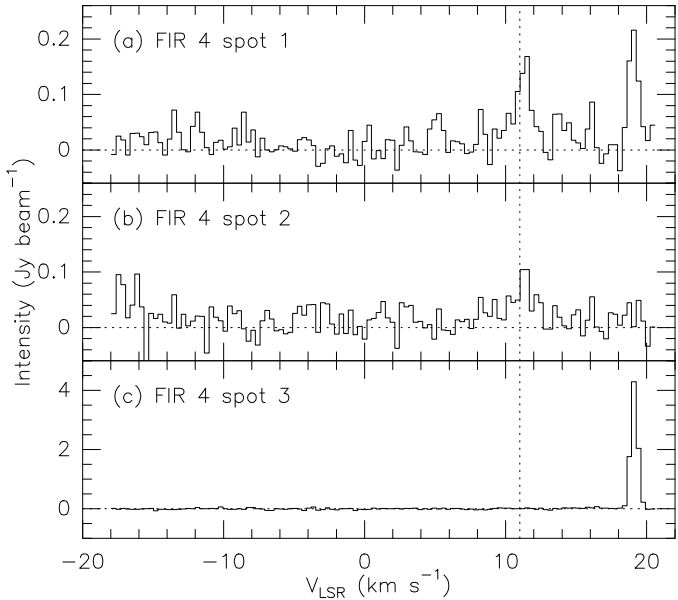}
\caption{
Spectra of the H$_2$O maser line in the FIR 4 region.
(a)
Spectrum of spot 1.
The peak at $\sim$19 km s$^{-1}$ is owing to the emission from spot 3.
(b)
Spectrum of spot 2.
(c)
Spectrum of spot 3.
Each spectrum was taken at a single pixel
at the peak position of the corresponding maser spot.
Vertical dotted line:
velocity of the ambient dense gas
($V_{\rm LSR}$ = 11.0 km s$^{-1}$; Schulz et al. 1991).}
\end{figure}

A Gaussian fit to spot 3 shows
that it is slightly larger than a point source,
with a deconvolved size of FWHM = 1\farcs0 $\times$ 0\farcs2.
The intensity distribution of spot 3
can be described as a combination of a strong point-source component
and a weak (below $\sim$10\% of the peak) extended component.
The weak component is elongated
in the north-south direction (P.A. = --10$^\circ$),
which caused the peak at $\sim$19 km s$^{-1}$
in the spectrum of spot 1 (Figure 6(a)).
Since this low-level elongation of spot 3
is symmetric around the intensity peak
and has a position angle exactly the same
as that of spot 3 with respect to the phase-tracking center,
it is probably an artifact of the imaging process,
not a real source structure.
It is probably owing to the position far from the phase-tracking center
and the velocity near the edge of the bandpass.

\section{DISCUSSION}

The positions of the radio continuum source and the H$_2$O maser spots
are shown in Figure 7,
together with other compact objects in the FIR 4 region.
Most of them are distributed along a straight line
in the northwest-southeast direction,
which suggests that they belong to a single protostar-outflow system.
The extended infrared feature in the 4.5 $\mu$m map (Figure 5),
extending from the infrared peak position toward the northwest,
can also be seen in the {\it Spitzer} maps
of other bands (3.6, 5.8, and 8.0 $\mu$m),
suggesting that it is seen in continuum emission.
The position and morphology of this feature
agree with those of the reflection nebulosity
seen in the $H$ and $K$ bands (Moore \& Chandler 1989).

\begin{figure}[!t]
\epsscale{1.0}
\plotone{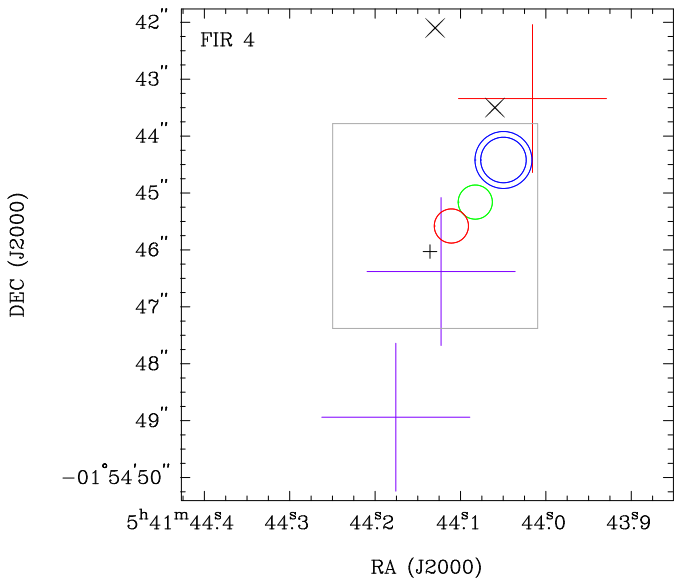}
\caption{
Compact objects in the FIR 4 region.
Small plus sign:
3.6 cm continuum source VLA 9 (the peak position in Figure 3).
Large plus signs:
H$_2$O maser spots (Figure 5).
Double circle:
near-IR source (2MASS Point Source Catalogue).
Open circles:
peak positions of the {\it Spitzer} maps
(green: 4.5 $\mu$m, red: 8.0 $\mu$m).
Gray square:
3 mm continuum source (Eisner \& Carpenter 2003).
Crosses:
X-ray sources (Skinner et al. 2003).
The size of each marker represents
the beam size of the corresponding observations for radio sources
or the pixel size for infrared and X-ray sources.}
\end{figure}

\subsection{FIR 4 Protostar}

The nature of the YSO embedded in the FIR 4 cloud core has been controversial
(Mezger et al. 1988, 1992; Moore \& Chandler 1989; Minier et al. 2003;
also see Section 4.2 of Choi et al. 2012a).
Close examinations of the available information reveal
that a major part of the confusion
stems from the infrared/millimeter source position issue.

Moore \& Chandler (1989) considered
that the point-like near-infrared (near-IR) source
represents the embedded YSO.
However, peak positions of the near-IR to mid-infrared (mid-IR) maps
do not coincide with each other (Figure 7).
They move toward the southeast with increasing wavelength,
along the outflow axis.
They tightly follow the straight line
defined by VLA 9 and the 2MASS source position.
This position shift suggests
that the detected near-IR emission comes from the scattered light
escaping through the outflow cavity.
The near-IR polarimetry is consistent with this interpretation
(Kandori et al. 2007).
This finding solves the extinction problem
pointed out by Mezger et al. (1992).
They argued that the expected extinction of the FIR 4 cloud core
is too high for any embedded YSO to be directly detectable in near-IR.
The central protostar is indeed invisible in near-IR.
Judging from the trend of the infrared position shift,
the true position of the protostar
may lie somewhere to the southeast of the 8.0 $\mu$m peak position.
The best candidate is the peak position of the radio thermal jet VLA 9.

High-resolution interferometric imaging in the millimeter continuum
can provide additional information on the position of the protostar.
Known from interferometric observations are the 3 mm source positions
reported by Chandler \& Carlstrom (1996) and Eisner \& Carpenter (2003).
Their positions of FIR 4 are inconsistent
and different by $\sim$2\farcs2 (comparable to their beam sizes).
A comparison of the positions of FIR 5 and 6 in the literature
(Wiesemeyer et al. 1997; Lai et al. 2002;
Alves et al. 2011; Choi et al. 2012b)
shows that the positions of Eisner \& Carpenter (2003)
are accurate to within the beam size,
but those of Chandler \& Carlstrom (1996)
are systematically shifted with respect to those of the other references.
The amount of shift is $\sim$3$''$ toward the northwest.
This astrometric problem caused some confusion in source identifications
(Minier et al. 2003; Skinner et al. 2003).
The FIR 4 position of Eisner \& Carpenter (2003)
agrees with VLA 9 within the beam size (Figure 7).
Therefore, VLA 9 is the best indicator
of the position of the FIR 4 protostar,
and there is no evidence for multiple YSOs in the FIR 4 dense core.

Setting aside the CH$_3$OH maser issue (Section 4.2),
most of the evidence indicates that FIR 4 contains a low-mass protostar.
Mezger et al. (1992) derived 10 $M_\odot$ for the mass of the cloud core
assuming that there is no internal heating and the core is cold ($\sim$19 K).
However, this assumption is inconsistent with recent findings.
Later studies showed that the temperature is higher
(50--80 K; Mangum et al. 1999; Watanabe \& Mitchell 2008).
Then the mass of the FIR 4 core may be much smaller: 0.8--1.7 $M_\odot$
(Chandler \& Carlstrom 1996; Visser et al. 1998; Watanabe \& Mitchell 2008).
Considering that FIR 4 is in an early stage of protostellar evolution,
the core mass seems to be too small
to form a high-mass/intermediate-mass star.

Moore \& Yamashita (1995) estimated
a bolometric luminosity of $\sim$25 $L_\odot$ for the YSO in FIR 4,
but this value is uncertain
because the far-infrared (far-IR) flux was not included in the calculation.
Lis et al. (1991) estimated a luminosity of 3--7 $L_\odot$
for the internal heating source
(or much smaller if the external radiation field is highly enhanced).
Therefore, the luminosity also indicates
that the FIR 4 core contains a low-mass protostar.

One of the major difficulties
in understanding the nature of the FIR 4 protostar is
that the nebulosity associated with the ionization front
and the infrared-bright stars in the nearby region
prevents accurate photometry in the far-IR band
where the spectral energy distribution peaks.
The luminosity from available data (discussed in the previous paragraph)
and the lack of an ultracompact/hypercompact H {\small II} region
firmly rule out the likelihood of FIR 4 containing a high-mass YSO.
The FIR 4 protostar is most likely a low-mass object,
but the possibility of an intermediate-mass (3--4 $M_\odot$) object
cannot be clearly ruled out.
Further investigations are needed to settle this issue.

\subsection{Methanol Maser Source}

One of the interesting features of FIR 4
is the CH$_3$OH 6.7 GHz maser emission (Minier et al. 2003).
This class II maser is considered to be exclusively associated
with massive star-forming regions (Menten 1991; Xu et al. 2008).
The CH$_3$OH maser spot of FIR 4
is one of the nearest class II maser sources known (Green et al. 2012)
and provides an excellent opportunity
to investigate the physical relation between YSOs and maser sources.

The CH$_3$OH maser source is closely associated
with VLA 9 and H$_2$O maser spot 1 (Figure 3).
The position marked in Figure 3
was taken from Figure 4 of Green et al. (2012),
which is more accurate than the one given by Minier et al. (2003).
The velocity of the CH$_3$OH maser line
($V_{\rm LSR}$ = 12.3 km s$^{-1}$; Minier et al. 2003; Green et al. 2012)
is redshifted by 1.3 km s$^{-1}$
relative to the systemic velocity (11.0 km s$^{-1}$; Schulz et al. 1991)
or by 1.0 km s$^{-1}$
relative to the velocity of the H$_2$O maser spot 1 (Table 3).
The widths of the CS and H$_2$CO lines from the FIR 4 dense core
are $\sim$2 km s$^{-1}$ (Schulz et al. 1991; Mangum et al. 1999).
The line profiles of H$_2$CO are nearly symmetric and have no line wing
(Mangum et al. 1999),
which suggests that the line width
mostly reflects the turbulent motion of dense molecular gas.
Therefore, it is unclear if the velocity shift of the CH$_3$OH maser
is related to the outflow, disk rotation,
or any other motion in the cloud core.

The angular separation
between the CH$_3$OH maser spot and the peak of the VLA 9
is $\sim$0\farcs21.
(The beam size of the CH$_3$OH line observations of Green et al. (2012)
is $\sim$0\farcs04.
The position accuracy of the VLA is usually better than 0\farcs1.)
The separation may not be completely accurate
because the source positions came from different observations.
Simultaneous observations in both the continuum and the CH$_3$OH maser line
are necessary to measure the separation more accurately.

Minier et al. (2003) argued
that FIR 4 contains a high-mass or intermediate-mass protostar,
based on the assumption
that the CH$_3$OH maser action may require
warm ($>$ 175 K) gas heated by the protostar.
They thought that the CH$_3$OH-rich gas in a low-mass star-forming region
exists far ($\gg$ 100 AU) from the central YSO
where the temperature is too low to drive a maser action.
The proximity of the CH$_3$OH maser to VLA 9
suggests that this argument needs to be reexamined.

The usual assumption is
that the maser spot is located in the heated envelope of a YSO
(Minier et al. 2003).
If the FIR 4 protostar is located exactly at the peak position of VLA 9,
the projected separation between the CH$_3$OH maser source and the protostar
is $\sim$90 AU.
The dust temperature at the CH$_3$OH maser-emitting region
can be estimated using Equation (2) of Motte \& Andr{\'e} (2001).
Assuming that the distance from the protostar to the maser spot
is $\sim$120 AU (considering the projection effect)
and that the protostellar luminosity is 3--25 $L_\odot$ (Section 4.1),
the dust temperature would be 40--70 K.
However, models of class II masers suggest
that mid-IR photons from warm (100--200 K) dust are necessary
for radiative pumping (De Buizer et al. 2000; Cragg et al. 2005).
Then the CH$_3$OH maser of FIR 4 suggests several possibilities.
(1) The class II maser action may operate
at temperatures lower than the theoretically expected range.
(2) The FIR 4 maser spot may have an extra source of heating,
such as the outflow or the external radiation field.
(3) The maser spot may be on the circumstellar disk
irradiated by photons coming directly from the protostar.
High-resolution ($\sim$0\farcs2 or better) observations in the dust continuum
may be helpful in understanding the relation
between the protostar and the CH$_3$OH maser source.
The flat radio spectrum, small mass, and low luminosity (Section 4.1)
rule out the existence of a high-mass YSO in the FIR 4 region,
and the CH$_3$OH maser must be excited by the FIR 4 low-mass protostar
regardless of the detailed heating and pumping mechanisms.

\subsection{FIR 4 Outflow}

The three H$_2$O maser spots are distributed along a straight line
in the northwest-southeast direction
with a P.A. of about \mbox{--23$^\circ$} (Figure 5).
This line passes through VLA 9
and agrees with the axis of the outflow in this region:
the northwestern lobe indicated by the reflection nebulosity
and the southeastern lobe traced by the CO line
(Moore \& Chandler 1989; Chandler \& Carlstrom 1996).
The unipolar CO outflow is redshifted (Chandler \& Carlstrom 1996).
The southeastern CO outflow may be on the far side of the FIR 4 core,
and the northwestern infrared nebulosity (outflow cavity)
may be on the near side.
Interestingly, the H$_2$O maser spot 3 in the northwest is redshifted,
which may reflect the velocity of the local shocked gas.

The centimeter continuum source VLA 9 (Figure 3)
is located close to H$_2$O maser spot 1.
VLA 9 is elongated,
and its major axis agrees with the direction of the outflow,
which suggests that VLA 9 is a radio thermal jet
at the base of the bipolar outflow.
With respect to the intensity peak,
the emission structure is more extended toward the northwest
than toward the opposite direction.

One of the X-ray sources reported by Skinner et al. (2003)
is located close to H$_2$O maser spot 3 (Figure 7).
The separation is $\sim$0\farcs7 or 300 AU.
The H$_2$O maser and the X-ray emission
are not necessarily coming from the same volume of gas,
but the northwestern outflow of FIR 4 may be responsible for both phenomena.
Only a few young protostars are known
to show the H$_2$O maser and X-ray emission together.
The YSOs in the NGC 2071 cluster (IRS 1/3 and VLA 1) are notable examples 
(Torrelles et al. 1998; Skinner et al. 2009; Trinidad et al. 2009).
High-resolution observations in the H$_2$O maser line
and deeper observations in the X-ray
are necessary to understand the physical relation
between the maser-producing shocked gas and the X-ray-emitting hot plasma.

\begin{figure*}[!t]
\epsscale{2.0}
\plottwo{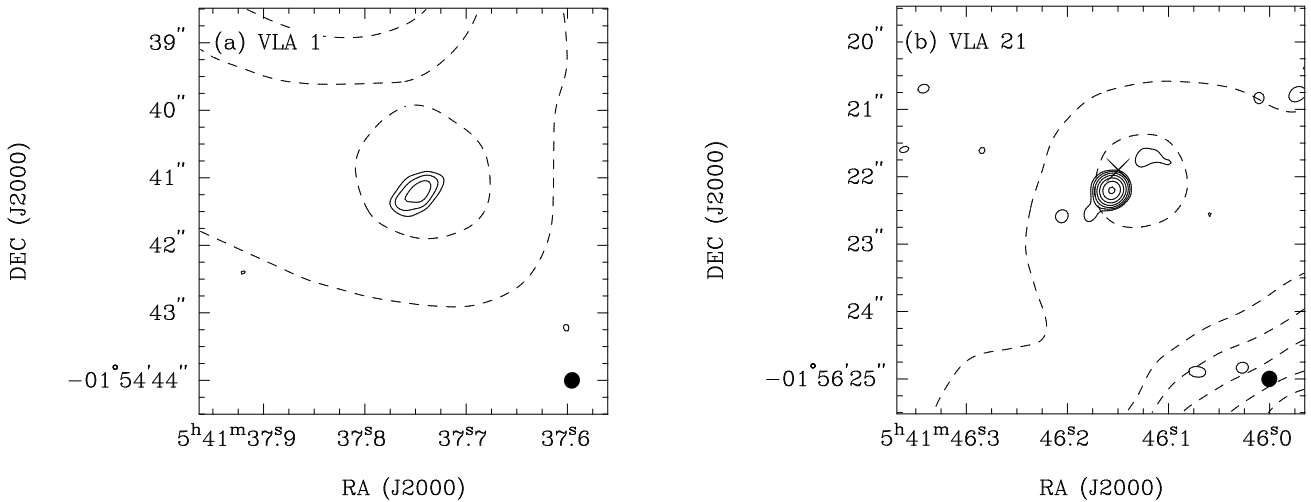}{f08b.eps}
\caption{
Maps of the 3.6 cm continuum emission around VLA 1 and VLA 21,
from the AR 465 data set.
The contour levels are
1, 2, 4, 8, 16, 32, and 64 $\times$ 0.05 mJy beam$^{-1}$.
The maps are corrected for the primary beam response.
Shown in the bottom right-hand corner is the synthesized beam:
FWHM = 0\farcs25 $\times$ 0\farcs23 with P.A. = --23$^\circ$.
The elongated shape of VLA 1 is owing to the bandwidth smearing.
Dashed contours:
maps of the 4.5 $\mu$m emission from the {\it Spitzer} data archive.
The contour levels are
1, 2, and 4 $\times$ 200 MJy sr$^{-1}$ for (a)
and 1, 2, 4, 8, 16, and 32 $\times$ 20 MJy sr$^{-1}$ for (b).
Cross:
X-ray source SGB 193 (Skinner et al. 2003).}
\end{figure*}

The four X-ray photons detected in the FIR 4 region
were hard ($\sim$5.9 keV; Skinner et al. 2003),
and the emission mechanism is difficult to explain.
Skinner et al. (2003) suggested
that the emission may be caused by stellar flares,
and additional possibilities (mainly relevant to NGC 2071 IRS 1)
are listed in Section 5.2 of Skinner et al. (2009).
The large separation ($\sim$2\farcs8)
between the X-ray source and the protostar (VLA 9)
seems to rule out all of these possibilities but the outflow model.
However, if the X-ray-emitting plasma is produced
by the outflow shocking on dense ambient gas,
the required shock speed is $\sim$2200 \mbox{km s$^{-1}$}
(see the discussion in Section 5.2.1 of Skinner et al. 2009).
This speed is unreasonably high for a YSO outflow.
Yet another model was proposed by L{\'o}pez-Santiago et al. (2013)
to explain the X-ray emission of HH 80.
In this model the hard X-ray emission is synchrotron radiation
produced by nonthermal processes in the magnetized jet.
It is unclear if this model applies to the case of FIR 4,
because there is no detection
of synchrotron radiation or strong magnetic fields.
Therefore, the origin of the hard X-ray remains a puzzle.

\section{NEARBY OBJECTS}

\subsection{Continuum Sources}

Rodr{\'\i}guez et al. (2003) detected
a cluster of 25 radio sources in the NGC 2024 region.
Based on a statistical consideration, they suggested
that practically all of these sources
are physically associated with NGC 2024.
Many of these radio sources are indeed associated with infrared sources
and other star formation activities such as outflows.
The nature of each source can be examined
with the spectrum of the centimeter continuum from ionized gas
(Reynolds 1986; Anglada et al. 1998).
Optically thin free--free emission (e.g., from a thermal radio jet)
shows a small spectral index ($\alpha \lesssim 1$).
Optically thick free--free emission
(e.g., from a high-density H {\small II} region)
shows a larger spectral index.
(At shorter wavelengths, the emission can become optically thin,
and the spectrum can become flat.)
Nonthermal emission (e.g., from a magnetic flare)
usually (but not always) shows a negative spectral index (G{\"u}del 2002).
Table 2 lists the spectral indices of 10 radio sources
that were detected in both 6.2 and 3.6 cm maps.
They can be classified into three categories
depending on the spectral slope.
The type of radio source is listed in the last column of Table 2.
(Also see the descriptions of several sources
in Rodr{\'\i}guez et al. 2003.)

\subsubsection{Positive-spectrum Sources}

Four sources (VLA 16, VLA 19, VLA 21, and VLA 25)
have relatively large spectral indices.
The centimeter continuua of these sources (except VLA 21) are
probably (partially) optically thick free-free emission.

VLA 16 is elongated in the northwest-southeast direction
(see Figure 5 of Rodr{\'\i}guez et al. 2003).
An elliptical Gaussian fit gives a deconvolved size
of FWHM = 0\farcs38 $\times$ 0\farcs15 with P.A. = --39$^\circ$.
The nature of VLA 16 is unclear.
It has an X-ray counterpart but remains undetected in the infrared
(Skinner et al. 2003).
Rodr{\'\i}guez et al. (2003) suggested
that VLA 16 forms a binary system with VLA 15 (IRS 2b).

VLA 19 (IRS 2) is the brightest radio/infrared source in this region.
The flux density increases steadily with the frequency,
and the spectral index in the 6.2--2.0 cm interval is $\sim$1.0 (Figure 4).
Kurtz et al. (1994) listed VLA 19 (G206.543--16.347)
as an ultracompact H {\small II} region containing a B2 star.
Lenorzer et al. (2004) suggested
that the radio emission comes from a stellar wind of an early-B-type star,
recombining at a radius of $\sim$100 AU.

VLA 21 is associated with an infrared source and an X-ray source (Figure 8).
Based on the high flux variability in a timescale of a few days,
Rodr{\'\i}guez et al. (2003) suggested
that the 3.6 cm continuum is gyrosynchrotron emission,
which seems to conflict with the positive spectral index.
(Considering the flux variability, however,
the uncertainty of the spectral index
can be much larger than the value given in Table 2.)
A possible explanation may be
that the radio flux of VLA 21 has both thermal and nonthermal components.
The quality of the multifrequency data
is not good enough to understand this problem.
Sensitive monitoring observations are necessary
to understand the variability of the flux and the spectral index.

\subsubsection{Flat-spectrum Sources}

Four sources (VLA 1, VLA 3, VLA 9, and VLA 24)
have relatively small spectral indices.
The centimeter continuum of these sources (except VLA 24) are
probably optically thin free-free emission.

Among the eight YSOs listed in Table 2,
only VLA 9 (FIR 4) is associated with a dense molecular cloud core
(Chandler \& Carlstrom 1996; Eisner \& Carpenter 2003).
VLA 9 is probably the youngest object among these YSOs.

Based on the circular polarization and mild flux variability
Rodr{\'\i}guez et al. (2003) suggested
that VLA 24 may be a young low-mass star showing gyrosynchrotron emission.
The positive spectral index suggests
that the radio flux of VLA 24 may also be a mixture
of thermal and nonthermal components.

\subsubsection{Negative-spectrum Sources}

Two sources (VLA 2 and VLA 8) have significantly negative spectral indices.
The centimeter continuum of these sources may be nonthermal emission.

VLA 8 is elongated in the north-south direction
(see Figure 3 of Rodr{\'\i}guez et al. 2003).
An elliptical Gaussian fit gives a deconvolved size
of FWHM = 0\farcs68 $\times$ 0\farcs31 with a P.A. = --6$^\circ$.
The spectral index in the 6.2--2.0 cm interval is about --1.5 (Figure 4).
Based on the morphology and the orientation of the elongation
Rodr{\'\i}guez et al. (2003) suggested
that VLA 8 may be an ionized proplyd.
However, the radio proplyd model conflicts
with the steeply decreasing radio spectrum.
Considering the extended structure, no day-scale variability,
and lack of infrared counterpart,
it is unlikely that VLA 8 is a magnetically active YSO
showing gyrosynchrotron emission.
VLA 8 is probably a background object such as an extragalactic radio jet.

\subsection{FIR 6n}

\begin{figure}[!b]
\epsscale{1.0}
\plotone{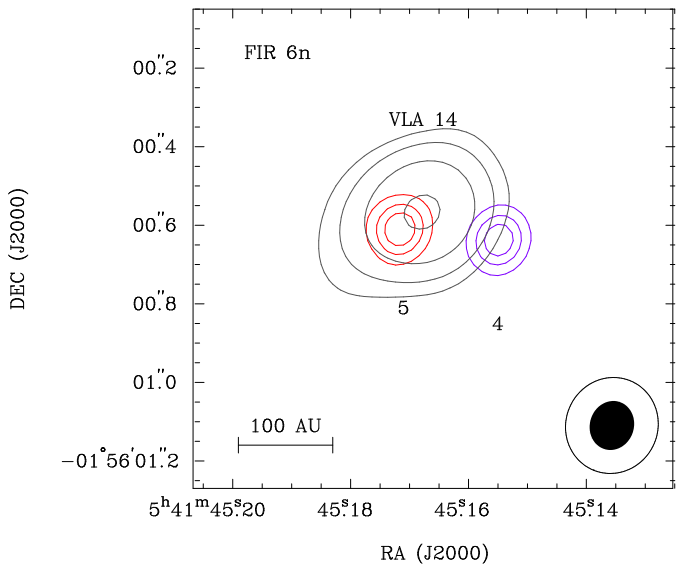}
\caption{
Maps of the H$_2$O maser and 3.6 cm continuum emission in the FIR 6n region.
The violet contours show the H$_2$O intensity
integrated over $V_{\rm LSR}$ = (9.3, 12.0) km s$^{-1}$,
and the contour levels are 0.25, 0.5, and 0.75 times the peak intensity
(8.88 Jy beam$^{-1}$ km s$^{-1}$).
The red contours show the H$_2$O intensity
integrated over (12.0, 23.8) km s$^{-1}$,
and the contour levels are 0.25, 0.5, and 0.75 times the peak intensity
(7.13 Jy beam$^{-1}$ km s$^{-1}$).
The maser spot numbers are labeled.
The gray contours show the 3.6 cm continuum intensity from the AR 465 data set,
and the contour levels are 1, 2, 4, and 8 $\times$ 0.03 mJy beam$^{-1}$.
The maps are corrected for the primary beam response.
Filled ellipse:
synthesized beam of the H$_2$O maps.
FWHM = 0\farcs13 $\times$ 0\farcs11 with P.A. = --17$^\circ$.
Open ellipse:
synthesized beam of the 3.6 cm continuum map.
FWHM = 0\farcs25 $\times$ 0\farcs23 with P.A. = --23$^\circ$.
The straight line in the bottom left-hand corner
corresponds to 100 AU at a distance of 415 pc.}
\end{figure}

Two H$_2$O maser spots associated with FIR 6n
were detected from the AM 780 observations:
one near the systemic velocity of the ambient cloud
($V_{\rm LSR}$ = 10.8 km s$^{-1}$; Choi et al. 2012a)
and the other at redshifted velocities (Figure 9).
The redshifted source (spot 5) shows three velocity components (Figure 10).
The H$_2$O maser sources detected previously
(Furuya et al. 2003; Choi et al. 2012b)
positionally coincides with spots 4/5,
but their beam sizes were too large to separate the two spots.

\begin{figure}[!t]
\epsscale{1.0}
\plotone{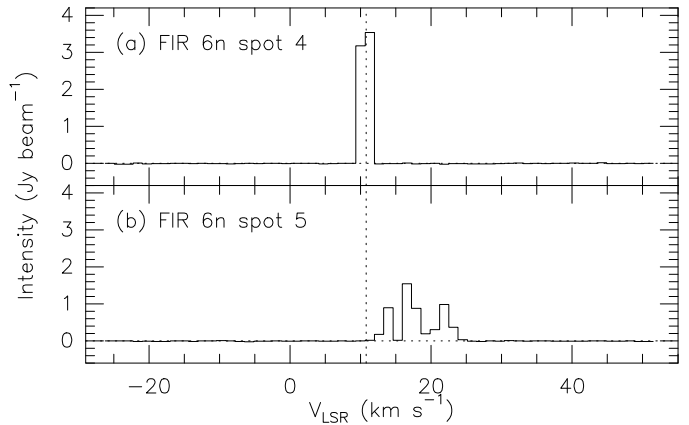}
\caption{
Spectra of the H$_2$O maser line in the FIR 6n region.
(a)
Spectrum of spot 4.
(b)
Spectrum of spot 5.
Vertical dotted line:
velocity of the ambient dense gas
($V_{\rm LSR}$ = 10.8 km s$^{-1}$; Choi et al. 2012a).}
\end{figure}

Figure 11 shows the positions of the H$_2$O maser spots
and other objects in the FIR 6 region.
The maser spots coincide with the millimeter continuum source FIR 6n.
The separation between the two spots
is much smaller than the angular resolutions
of the millimeter continuum maps available
(Lai et al. 2002; Alves et al. 2011; Choi et al. 2012b).
The centimeter continuum source VLA 14 (Rodr{\'\i}guez et al. 2003)
coincides with H$_2$O maser spot 5 (Figure 9).
VLA 14 is unresolved.
The peak position and flux density given by Rodr{\'\i}guez et al. (2003)
are consistent with those from our analysis.

\begin{figure}[!b]
\epsscale{1.0}
\plotone{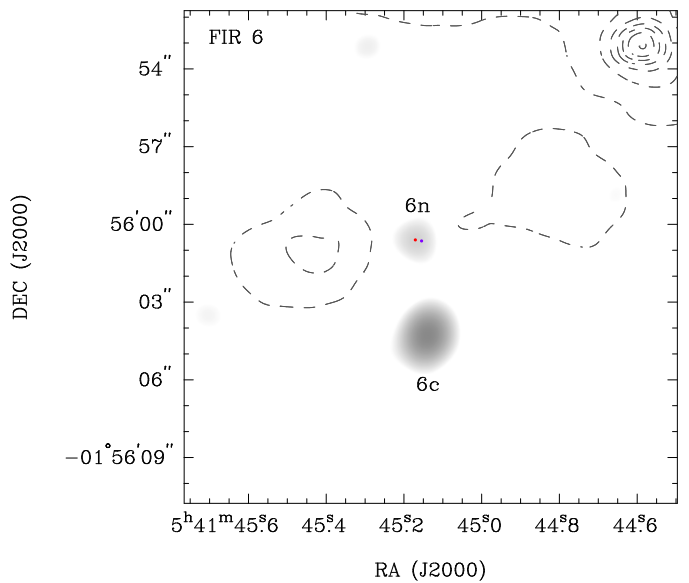}
\caption{
Positions of the compact objects in the FIR 6 region.
Filled circles:
H$_2$O maser spots (Figure 9).
Grayscale:
map of the 6.9 mm continuum emission (see Figure 3 of Choi et al. 2012b).
Dashed contours:
map of the 4.5 $\mu$m emission from the {\it Spitzer} data archive.
The contour levels are 2, 3, 4, 5, 6, 7, and 8 $\times$ 13 MJy sr$^{-1}$.
The emission in the northwestern corner of the map
comes from an infrared star.}
\end{figure}

The two maser spots are separated in the east-west direction (Figure 9).
The separation (0\farcs24) corresponds to 100 AU,
and the relative position angle is 81$^\circ$.
This value is similar to the position angle
of the 30$''$ scale bipolar outflow traced by the CS line
and the high-velocity CO $J$ = 1 $\rightarrow$ 0 emission
($\sim$75$^\circ$; Chandler \& Carlstrom 1996).
The extended infrared feature in the 4.5 $\mu$m map (Figure 11)
shows two emission peaks, one in the east and the other in the west,
and seems to be related to the outflow.
The 4.5 $\mu$m emission feature
is not obvious in the {\it Spitzer} maps of the other bands,
suggesting that it probably comes from line emission in the 4.5 $\mu$m band.
The separation between the 4.5 $\mu$m peaks is $\sim$10$''$,
and the relative position angle is $\sim$106$^\circ$,
which is different from that of the CS/CO outflow by $\sim$30$^\circ$.
Alves et al. (2011) showed
that the outflow traced by the CO $J$ = 3 $\rightarrow$ 2 line
has a complicated morphology
and suggested that the outflows of FIR 6n and FIR 5 may be interacting.

FIR 6n is the most active H$_2$O maser source
in the whole NGC 2024 molecular ridge region (Choi et al. 2012a).
The nature of the YSO in FIR 6n is unclear
because, in the millimeter/infrared wavelengths,
FIR 6n is often either undetectable or affected by FIR 6c.
Considering that FIR 6n drives a strong outflow
and is invisible in the near-IR,
the FIR 6n cloud core may contain a low-mass protostar.
The nature of the stronger millimeter/submillimeter source FIR 6c
is even more unclear.
There is no molecular outflow associated with FIR 6c,
but a redshifted H$_2$O maser was detected (Choi et al. 2012b).
FIR 6c was undetected in the 3.6 cm continuum.
The spectral energy distribution of FIR 6c rises steeply around 6.9 mm
and is flat around 1 mm,
which suggests that FIR 6c probably contains
a hypercompact H {\small II} region (Choi et al. 2012b).
If so, the central object may be an early-B type star.
High-resolution imaging in the millimeter continuum
is necessary to understand this peculiar object.

\section{SUMMARY}

The star formation activities in the NGC 2024 region
were investigated by analyzing several archival data sets from the VLA.
The nature of centimeter continuum sources was examined
using a data set of multifrequency observations.
The source positions and structures were inspected
using the data set of 3.6 cm continuum observations
previously published by Rodr{\'\i}guez et al. (2003).
Outflow activities were studied
using several data sets of H$_2$O maser observations.
Careful examinations of the positions of compact objects in the FIR 4 region
provide a simple and comprehensible picture
of the star formation activities around the FIR 4 protostar.
The main results are summarized as follows.

1.
The centimeter continuum source VLA 9, the source associated with FIR 4,
shows a flat spectrum in the 6.2--3.6 cm interval.
VLA 9 is elongated in the direction consistent with the FIR 4 outflow.
VLA 9 is a radio thermal jet
and the best indicator of the protostellar position.
The flat spectrum rules out
the possibility of a high-mass YSO in the FIR 4 region.
Previously known mass and luminosity estimates suggest
that the FIR 4 protostar is most likely a low-mass object.

2.
The CH$_3$OH 6.7 GHz maser source in FIR 4 is located close to VLA 9.
The separation between them is $\sim$0\farcs2.
They must have a close physical relation,
but the detailed amplification mechanism of the maser is unclear.
The CH$_3$OH maser of FIR 4 demonstrates
that the class II maser phenomenon
does not necessarily require the presence of a high-mass YSO.

3.
The H$_2$O maser spots of FIR 4
are distributed along the FIR 4 outflow axis.
The strongest spot is associated with an X-ray source.
The FIR 4 outflow is a rare example of protostellar outflows
showing the H$_2$O maser and X-ray emission together.
Further studies are needed
to understand the origin of the hard X-ray emission.

4.
The properties of other radio sources in the field of view were investigated.
VLA 1, VLA 3, and VLA 25 (IRS 4) may be radio thermal jets driven by YSOs.
VLA 2, VLA 21, and VLA 24 may be magnetically active YSOs.
VLA 8 may be a background object emitting nonthermal radiation.
VLA 16 is a radio source of unknown nature.
VLA 19 (IRS 2) may be an early-B-type star.

5.
The two H$_2$O maser spots of VLA 14 (FIR 6n)
are distributed along an east-west line
that is consistent with the axis of the FIR 6n molecular outflow.
The FIR 6n cloud core may contain a low-mass protostar.

\acknowledgements

We thank Karl M. Menten for helpful discussions.
M. C. was supported by the Core Research Program
of the National Research Foundation of Korea (NRF)
funded by the Ministry of Science, ICT and Future Planning
of the Korean government (grant No. NRF-2011-0015816).
J.-E. L. was supported by the Basic Science Research Program through NRF
funded by the Ministry of Education of the Korean government
(grant No. NRF-2012R1A1A2044689).
NRAO is a facility of the National Science Foundation
operated under cooperative agreement by Associated Universities, Inc.
This work is based in part on observations
made with the {\it Spitzer Space Telescope},
which is operated by the Jet Propulsion Laboratory,
California Institute of Technology, under a contract with NASA.

\end{document}